\documentclass[5p]{elsarticle}

\usepackage{graphicx}

\usepackage{amssymb}

\usepackage[nodots]{numcompress}

 \biboptions{sort&compress}

\begin{document}

\begin{frontmatter}



\title{Investigating transition state resonances in the time domain\\
by means of Bohmian mechanics: The F+HD reaction}


\author{A. S. Sanz}

\author{D. L\'opez-Dur\'an}

\author{T. Gonz\'alez-Lezana}

\address{Instituto de F\'{\i}sica Fundamental - CSIC, Serrano 123,
28006 Madrid, Spain}

\begin{abstract}
In this work, we investigate the existence of transition state
resonances on atom-diatom reactive collisions from a time-dependent
perspective, stressing the role of quantum trajectories as a tool to
analyze this phenomenon.
As it is shown, when one focusses on the quantum probability current
density, new dynamical information about the reactive process can be
extracted.
In order to detect the effects of the different rotational
populations and their dynamics/coherences, we have considered a
reduced two-dimensional dynamics obtained from the evolution of a
full three-dimensional quantum time-dependent wave packet associated
with a particular angle.
This reduction procedure provides us with information about the
entanglement between the radial degrees of freedom $(r,R)$ and the
angular one $(\gamma)$, which can be considered as describing an
environment.
The combined approach here proposed has been applied to study the
F+HD reaction, for which the FH+D product channel exhibits a
resonance-mediated dynamics.
\end{abstract}


\end{frontmatter}



\section{Introduction}
\label{sec1}

The analysis in the time domain of processes and phenomena that appear
in chemical reactions constitutes nowadays a strong field of research
both experimentally and theoretically.
From the experimental point of view,
the femtosecond techniques developed in the 1990s have
allowed us to monitor the process step by step, from reactants to
products, with an incredibly high resolution.
This has lead to an impressive amount of theoretical work based on
wave packet (WP) propagation techniques.
In general, these studies are aimed at computing very long time series,
from which energy/frequency spectra are obtained at a high resolution
(the longer the timescale considered, the finer the features one can
define in the corresponding energy/frequency spectrum), which are later
on compared with the experiment.
Within this scenario, little attention is eventually paid to the
evolution of the WP itself, apart from the frame-to-frame picture of
the process.
Not much information is usually extracted from the corresponding
probability density, except for where the latter displays maxima or,
on the contrary, presents regions of voids (nodal regions).

In order to obtain an alternative description of the dynamics, one can also
analyze the quantum probability current density.
Usually this quantity is only considered as a quantum flux to determine
how much of the probability density flows into or out a certain region
of the configuration space defining the system of interest.
It also gives us an idea
on how the former evolves in time, just as currents or streams allow us
to visualize how a river flows along its bed.
This additional information has nevertheless not been much exploited in
the literature, except for the study of magnetic properties (because of
the needs imposed by the nature of the magnetic field, which is
sourceless and therefore has to be considered in hydrodynamic terms).
In this sense, Bohmian mechanics \cite{bohm,holland-bk}, a reformulation
of standard quantum mechanics in terms of trajectories or, equivalently,
streamlines, results of much help.
This approach enables us to take the analogy
of the river further away: the trajectories behave like the paths pursued
by some tracer particles that travel with the stream, thus giving us
a precise description of the flow dynamics.
This information cannot be obtained directly from the quantum
probability current density.
It provides us with an insight at a local level of the stream, but does
not say anything about how a particular point-like particle evolves
when it starts from a certain initial position in the corresponding
configuration space.
Within this approach, quantum trajectories (QTs) or streamlines are
mainly obtained through any of two approaches \cite{wyatt-bk}:
synthetic \cite{wyatt1,wyatt2,wyatt-bk} and analytic
\cite{sanz-ssr,sanz-pr}.
Within the former, QTs are computed after solving simultaneously the
quantum Hamilton-Jacobi equation and the continuity one, while in the
latter (used in this work) one starts from Schr\"odinger's equation and
then the QTs are obtained from the phase of the wave function.

The use of Bohmian mechanics with interpretational purposes can be
traced back to 1926, when Madelung \cite{madelung} tackled the issue of
the interpretation of the wave function (in fashion at the time) by
reformulating quantum mechanics in hydrodynamic terms.
These ideas underwent a rebirth in the 1970s through the works of
Bialynicki-Birula \cite{birula1,birula2} and Hirschfelder
\cite{hirsch1,hirsch2,hirsch3,hirsch4}.
Within this hydrodynamical framework, treating the probability density
as a quantum fluid, the chemical reactivity of collinear reactions was
formerly studied by the end of the 1960s and beginning of the 1970s by
McCullough and Wyatt \cite{MC-W1,MC-W2,MC-W3}.
This can be considered the starting point for the use of this theory
in chemical physics, first at an analytic level and then at
a synthetic one by Lopreore and Wyatt \cite{wyatt1,wyatt2}.
In this latter case, it constituted the source from which a series
of numerical algorithms aimed at obtaining the evolution of quantum
systems were developed, thus leading to the so-called QT methods
\cite{wyatt-bk}.
In the same direction, more recently Sanz {\it et al.}\ have analyzed
\cite{sanz-bofill1,sanz-bofill2} the dynamics associated with a
prototype of reactants-to-products
reaction described by the so-called M\"uller-Brown potential energy
surface (PES) \cite{muellerbrown}.
On the other hand, the QT methodology has also been exploited by
different authors \cite{as:bader,as:gomes-1,as:gomes-2,as:lazzeretti-1,%
as:lazzeretti-2,as:lazzeretti-3,as:lazzeretti-4,as:lazzeretti-5,%
as:lazzeretti-6,as:lazzeretti-7,as:lazzeretti-8} to understand the
magnetic properties of molecules within a framework that encompasses
electronic structure and topology.

Different theoretical methods have been developed in the literature to
analyze, to explain and to understand the formation of transition state
resonances.
One of the reasons for such an interest relies on the crucial role played
by these processes as intermediate stages in chemical reaction dynamics
and aggregation processes to form atomic clusters.
In this sense, the F+HD reaction constitutes a paradigmatic system.
This system has been actively investigated in the last decade due to the
the observation through molecular beam experiments of a resonance for the
reactive channel F+HD $\to$ HF+D \cite{SSMLDL:JCP00,SSMLDL:PRL00}.
In the differential cross section, this dynamical feature manifests as
a sharp variation of the peaks along the forward-backward scattering
directions and, in the corresponding integral cross section, as a
step-like profile.
Quantum mechanical (QM) calculations on the Stark-Werner (SW)
{\it ab initio} PES \cite{SW:JCP96} showed that those observations are
consistent with a resonance-mediated dynamics of the reaction at a
collision energy near 0.5~kcal$\cdot$mol$^{-1}$ (0.022~eV).
In particular, the existence of a resonance with $(v_{\rm FH}=3,
v_{\rm HD}=0)$ quantum numbers for the F--H and H--D stretching modes,
respectively, located at the transition state region was theoretically
proved \cite{SSMLDL:JCP00,SSMLDL:PRL00}.
On the contrary, the dynamics of the other product channel, F+HD $\to$
FD+H, was found not to display evidences of similar features.

The specific location of this resonant feature, just above the entrance
channel, minimizes the number of partial waves contributing to the
integral cross section.
This also explains why the peak observed is capable to survive to
the averaging introduced when all such partial waves are taken into
account.
In addition, the resonance lies just below the threshold for reaction predicted by
classical approaches, as it has been shown by means of quasi-classical
trajectory calculations \cite{ABHSSW:JCP95,ABHSSTW:CPL96}.
Although the maximum peak associated with the resonance was not
reproduced in those studies, the rotational motion was found to
play a significant role in the overall reaction mechanism.
Thus, in order to provide a correct description of the dynamics of the
F+HD $\to$ FH+D reaction, a complete three-dimensional (3D) QM approach
is required \cite{SSMLDL:JCP00}.

In this work we have carried out a fully converged QM study for a zero
total angular momentum ($J=0)$ by means of a 3D time-dependent wave
packet (TDWP) method.
In an attempt to further investigate the interpretive capabilities of
trajectory-based approaches for this kind of reactive processes, a
QT method has been employed in combination with the propagated WP.
Thus, besides the 3D WP propagation, projections onto specific values
of the angular coordinate ($\gamma$) have been considered to run an
alternative QT study of the process.
The reason to consider the projection-based analysis presented here
arises as a way to avoid the complexity of the full 3D QT dynamics at
a first stage, attacking the study of the resonance by only considering
the QT propagation within the two-dimensional (2D) radial subspace.
In spite of its limitations, this 2D dynamical study is shown to
be insightful when compared to the corresponding complete 3D TDWP
calculations.
More specifically, this Bohmian analysis allows us to determine how
the energy/probability associated with a particular angle flows and
can be related to the resonance.

This work is organized as follows.
In Section \ref{sec2} the theoretical details of the TDWP and QT
approaches are explained.
Results are shown and discussed in Section \ref{sec3}.
Finally, the main conclusions extracted from this work are summarized
in Section \ref{sec4}.


\section{Theoretical framework}
\label{sec2}


\subsection{The WP procedure}
\label{sec21}

The TDWP method has been described before \cite{GRM:JCP04}, so here we
will only describe the main details involved.
Accordingly, consider the total Hamiltonian, $H = T + V$, is described
in mass-scaled Jacobi coordinates.
With this choice, the kinetic energy operator, $T$, for $J=0$ reads
as
\begin{equation}
 T = - \frac{\hbar^2}{2\mu} \left[ \frac{\partial^2}{\partial R^2}
   + \frac{\partial^2}{\partial r^2} + \frac{{\bf l}^2}{2\mu R^2}
   + \frac{{\bf j}^2}{2\mu r^2} \right] ,
 \label{kinetic}
\end{equation}
where ${\bf l}$ and ${\bf j}$ are the orbital angular and diatomic
rotational angular momentum, respectively, and $V(R,r,\hat{\bf R}
\cdot \hat{\bf r})$ is the potential energy function, which in this
work corresponds to the SW PES \cite{SW:JCP96}.

The WP time propagation can be formally expressed as
\begin{equation}
 |\phi_i(t)\rangle = e^{-i(H-i\epsilon)t/\hbar} |\phi_i(0)\rangle .
 \label{propag}
\end{equation}
Here, this propagation is performed using the symmetric split-operator
formula \cite{FF:JCP83},
\begin{equation}
 e^{-i(H-i \epsilon) \tau / \hbar} = e^{-i V \tau / 2 \hbar}
 e^{-i(T - i\epsilon) \tau / \hbar}
 e^{-iV \tau / 2 \hbar} + O(\tau^3) .
 \label{split}
\end{equation}
In Eq.~(\ref{propag}), the initial-state selected WP, $\phi_i(0)$, is
assumed to be a product state with the form
\begin{equation}
 \phi_i(0) = G(R) \phi_{vj}(r) Y_{lj}^{JM}(\hat{\bf R},\hat{\bf r}),
 \label{initwp}
\end{equation}
where $G(R)$ is the localized translational wave function,
$\phi_{vj}(r)$ is the rovibrational wave function for the
selected initial HD state, and
\begin{equation}
 Y_{lj}^{JM}(\hat{\bf R}, \hat{\bf r}) =
  \sum_{m_l, m_j} \langle l m_l, j m_j | J M \rangle
   Y_{l m_l}(\hat{\bf R}) Y_{j m_j} (\hat{\bf r})
 \label{harmonic}
\end{equation}
is a bipolar harmonic.
The translational wave function $G(R)$ is given by a Gaussian WP of
width $\Delta R_0$ and average incident momentum $\hbar k_0$, i.e.,
\begin{equation}
 G(R) = e^{-(R-R_0)^2/2 \Delta R^2_0 - ik_0 R} .
 \label{gauss}
\end{equation}

As it is explained in Ref.~\cite{GRM:JCP04}, a discrete variable
representation is employed to describe the radial $R$ and $r$
coordinates, while a basis set of bipolar harmonics is chosen for
the angular degree of freedom, $\gamma$.
For asymmetric reactions, as in the case considered here, $\gamma$
defines separate regions that correspond to two different product
arrangements \cite{GRM:JCP04}.
In order to avoid numerical reflections as the WP propagates along the
different channels associated with the reaction, absorption of the
corresponding outgoing probability fluxes is also considered.
In this regard, for the absorbing potential $-i \epsilon(R,r) =
i \epsilon_R(R) - i \epsilon_r(r)$ in Eq.~(\ref{propag}), we have
used the expression proposed in Ref.~\cite{GRM:JCP04} for the potential
originally developed by Manolopoulos \cite{M:JCP02}.
This is applied at $r_1$ and $R_1$, specific values at the end of the
corresponding grids.

The values for the parameters used in the TDWP calculation throughout
this work, such as parameters for the initial WP (the center $R_0$ of
the Gaussian function $G(R)$, the incident wave vector $k_0$, and the
width $\Delta R_0$), the time step ($\tau$), the total propagation
time ($t_{tot}$), the maximum value of diatom rotational states
($j_{max}$), the step for the radial grids ($\delta r$ and $\delta R$)
and the position of the absorbing potential for the radial coordinates
($R_1$ and $r_1$) are given in Table \ref{tabTDWP}.


\subsection{A brief account on Bohmian mechanics}
\label{sec22}

Consider a system of mass $m$ in Cartesian coordinates.
In Bohmian mechanics, the wave function $\Psi$ associated with this
system provides us with dynamical information about it at any point on
configuration space and any time.
More specifically, this information is encoded in the phase of $\Psi$,
as infers from the transformation relation
\begin{equation}
 \Psi({\bf r},t) = \rho^{1/2}({\bf r},t) e^{iS({\bf r},t)/\hbar} ,
 \label{e2}
\end{equation}
where $\rho$ and $S$ are the probability density and phase of
$\Psi$, respectively, both being real-valued functions.
From Eq.~(\ref{e2}), Schr\"odinger's equation,
\begin{equation}
 i\hbar\ \frac{\partial \Psi}{\partial t} =
  \left( - \frac{\hbar^2}{2m}\ \nabla^2 + V \right) \Psi ,
 \label{e1}
\end{equation}
is recast as a system of two real coupled equations,
\begin{eqnarray}
 \frac{\partial \rho}{\partial t} & + &
  \nabla \cdot \left( \rho\ \frac{\nabla S}{m} \right) = 0 ,
 \label{e3} \\
 \frac{\partial S}{\partial t} & + & \frac{(\nabla S)^2}{2m} +
  V + Q = 0 ,
 \label{e4}
\end{eqnarray}
where
\begin{eqnarray}
 Q \equiv - \frac{\hbar^2}{2m} \frac{\nabla^2 \rho^{1/2}}{\rho^{1/2}}
 = \frac{\hbar^2}{4m}
   \left[ \frac{1}{2} \left( \frac{\nabla \rho}{\rho} \right)^2
   - \frac{\nabla^2 \rho}{\rho} \right]
 \label{e6}
\end{eqnarray}
is the so-called quantum potential.
Equation~(\ref{e3}) is the continuity equation, which rules the
ensemble dynamics of a swarm of trajectories with initial positions
distributed according to $\rho_0$; Eq.~(\ref{e4}) is the quantum
Hamilton-Jacobi equation, which describes the phase field evolution
ruling the motion of quantum particles through the motion equation
\begin{equation}
 {\bf v} = \dot{\bf r} = \frac{\nabla S}{m} .
 \label{e5}
\end{equation}
The coupling between Eqs.~(\ref{e3}) and (\ref{e4}) through $Q$ (or,
equivalently, $\rho$) is the reason why quantum (Bohmian) dynamics is
very different from its classical counterpart, where both equations are
only coupled through $S$.
This therefore constitutes a very important difference between the
typical calculations based on classical trajectories, commonly used in
chemical reactivity, although they cannot reproduce quantum features
such as tunneling or interference, and simulations based on Bohmian
trajectories.
In other words, this coupling is precisely the way how the wave function
guides the motion of the particle and hence allows QTs to display true
quantum features.

\begin{table}[t]
 \caption{Parameters used in the TDWP calculations carried out here
  (see text for details).}
 \label{tabTDWP}\tabcolsep=20pt
 \begin{center}
  \begin{tabular}{cc}
  \hline \hline
   Parameter & Value \\
   \hline
   $R_0$ (bohr) & 9.6 \\
   $k_0$ (bohr$^{-1}$) & 0.42 \\
   $\Delta R_0$ (bohr) & 0.09 \\
   $\tau$ (fs) & 0.016 \\
   $t_{tot}$ (fs) & 165 \\
   $j_{max}$ & 95 \\
   $\delta R$ (bohr) & 0.061 \\
   $\delta r$ (bohr) & 0.052 \\
   $R_1$ (bohr) & 11.0 \\
   $r_1$ (bohr) & 8.0 \\
  \hline \hline
  \end{tabular}
 \end{center}
\end{table}

If instead of trajectories, we are more interested in looking at the
quantum system dynamics as the evolution of a quantum fluid, i.e., in a
quantum hydrodynamical view, the magnitudes of interest will be the
probability density, $\rho = \Psi^*\Psi$, and the probability current
density, ${\bf J} = \rho {\bf v} = \rho (\nabla S/m)$, related
through the continuity equation (\ref{e3}), as
\begin{equation}
 \frac{\partial \rho}{\partial t} = - \nabla \cdot {\bf J} .
 \label{e9}
\end{equation}
Hence, instead of talking about trajectories, one speaks about the
conservation of the quantum flow, where the analog of Eq.~(\ref{e4})
for ${\bf v}$ is a quantum Euler or Navier-Stokes equation and the
solutions of Eq.~(\ref{e5}) are regarded as fluid streamlines rather
than trajectories.
These streamlines are obtained by integrating
\begin{equation}
 {\bf v} = \dot{\bf r} = \frac{{\bf J}}{\rho} ,
 \label{e10}
\end{equation}
which is formally equivalent to Eq.~(\ref{e5}).
Accordingly, these lines would follow the flow described by the quantum
(probabilistic) fluid which describes the system.

In this work, we are going to explore the 2D restricted dynamics
associated with the subspace ($R,r$) describing the F+HD $\to$
FH+D/FD+H reaction.
Assuming the last two terms of the kinetic operator given in
Eq.~(\ref{kinetic}) can be considered as a part of some effective
potential operator, the motion equations describing the evolution
of the corresponding reduced QTs will be
\begin{eqnarray}
 \dot{R} & = & \frac{1}{\mu} \frac{\partial S_\gamma}{\partial R} ,
 \label{dotR} \\
 \dot{r} & = & \frac{1}{\mu} \frac{\partial S_\gamma}{\partial r} ,
 \label{dotr}
\end{eqnarray}
where $S_\gamma$ is the phase associated with the 2D time-de\-pen\-dent
reduced wave function $\phi_\gamma (R,r;t)$.
This wave function is obtained by projecting of the total 3D wave
function written in Eq.~(\ref{propag}) onto a given angle $\gamma$.
In principle, this projected wave function obeys a continuous, smooth
evolution in the reduced subspace and, therefore, one also expects the
corresponding reduced QTs to display a continuous evolution.
Note that this is somehow equivalent to describe the dynamics
associated with an effective Hamiltonian
\begin{equation}
 H_\gamma = - \frac{\hbar^2}{2\mu} \frac{\partial^2}{\partial R^2}
            - \frac{\hbar^2}{2\mu} \frac{\partial^2}{\partial r^2}
            + V_{\rm eff}(R,r;\gamma) .
 \label{effHam}
\end{equation}
Therefore, although this reduced dynamics does not provide a full
picture of the 3D system, as we shall see it results very insightful
to understand the mechanism of the resonance process in a simplified
manner.

It is interesting to stress here that, due to its dependence on all
the remaining values of the angular coordinate, the evolution under
the Hamiltonian $H_\gamma$ can be understood as affected by a sort of
environment coordinate.
No reduction of the latter by tracing over the associated full 3D
density matrix or by means of related phenomenological reduced model
\cite{borondo1,borondo2} is used.


\subsection{Quantum trajectories for a Gaussian WP}
\label{sec23}

One of the relevant aspects of the process concerns the preparation of
the initial state, for it will determine the subsequent evolution of
the chemical species.
In order to render some preliminary light on this issue, here we
are going to briefly analyze the evolution of our 2D projected WP.
The wave function describing our system's state is given by the product
of a Gaussian WP in the $R$-coordinate and the ground state
of the asymptotic potential along the $r$-coordinate.
This means that, until the wave function is not well inside the
interaction region, we will essentially have a wave function which
is a free expanding and onwards propagating Gaussian with basically
the same width along $r$.
Thus, we can simplify the analysis by only considering the evolution
of a Gaussian wave function evolving along the $R$-coordinate for
times lesser than the time $t_{int}$ at which the interaction is
strong enough as to couple the two radial degrees of freedom and the
(rotational) angular ones.
In this sense, consider $t \ll t_{int}$.
The evolution of (\ref{gauss}), when normalized, is then approximately
described by
\begin{equation}
 G(R,t) = A_t e^{-(R - R_{cl})^2/4\tilde{\sigma}_t\sigma_0
   + ik_0 (R - R_{cl}) + iEt/\hbar} .
 \label{eq11}
\end{equation}
where $A_t = (2\pi\tilde{\sigma}_t^2)^{-1/4}$ and $\Delta R_0 =
\sqrt{2} \sigma_0$.
This WP moves along the classical trajectory $R_{cl} =
v_0 t$ (for simplicity, we have assumed a zero value for
the initial $R_{cl,0} = 0$) and its
(complex) spreading with time goes as $\tilde{\sigma}_t = \sigma_0
(1 + i\hbar t/2m\sigma_0^2)$.
The initial velocity by $v_0 = \hbar k_0/\mu = \langle \hat{p}/\mu
\rangle$ and the initial energy by $E = \hbar^2 k_0^2/2\mu$.
The phase (Bohmian action) corresponding to (\ref{eq11}) then reads as
\begin{eqnarray}
 S_G(R,t) & = & - \frac{\hbar}{2} \ \! \arctan \left(\frac{\hbar t}
  {2\mu\sigma_0^2} \right) + E t \nonumber \\
 & & + \hbar k_0 R
     + \frac{\hbar^2 t}{8\mu\sigma_0^2\sigma_t^2} \ \! (R - R_{cl})^2 ,
 \label{eq13b}
\end{eqnarray}
with
\begin{equation}
 \sigma_t = |\tilde{\sigma}_t| =
  \sigma_0 \sqrt{1 + \left( \frac{\hbar t}{2\mu\sigma_0^2} \right)^2}
 \label{e12}
\end{equation}
being the time-dependent real spreading.

The role of the quantum phase in the reaction dy\-na\-mics is very
important and becomes very apparent through the analysis of QTs.
An analysis of the early stages of the reaction results very insightful
in order to understand the further time-evolution.
As said above, these stages are well described by the Gaussian WP
(\ref{eq11}).
Thus, introducing (\ref{eq13b}) into (\ref{dotR}) and then integrating
in time, we find
\begin{equation}
 R_q(t) = R_{cl}(t) + \frac{\sigma_t}{\sigma_0}\ R_{q,0} ,
 \label{e13}
\end{equation}
where $R_{q,0}$ is the initial position of the corresponding trajectory.
Notice that, in order to get this result, we have assumed that
$\phi_\gamma(R,r;t)$ can be considered separable in $G(R,t)$ and
another function depending on $r$ at these early stages of the
evolution, which also means that
\begin{equation}
 r_q(t) \approx r_{q,0} ,
 \label{e13r}
\end{equation}
i.e., at short times QTs only evolve along the $R$-coor\-di\-na\-te,
their position remaining essentially constant along the $r$-coordinate.

In Eq.~(\ref{e13}) we clearly distinguish the two contributions ruling
the time dependence of $G(R)$.
First, a classical drift which makes any QT to move alongside the
corresponding classical path.
This could be therefore regarded as a Bohmian classicality criterion,
to some extent related to Ehrenfest's theorem (see below).
The second contribution is a quantum fluid drift,
which comes from the own (quantum) nature of the evolution of
$\phi_\gamma$ and describes the (free) expansion of the WP
with time.
Accordingly, QTs will separate among themselves at a nonuniform
accelerated rate or velocity,
\begin{equation}
 \frac{dR_q}{dt} = \frac{\sigma_0}{\sigma_t}
   \frac{t}{\tau_q^2}\ R_{q,0} ,
 \label{e14}
\end{equation}
with $\tau_q \equiv 2\mu\sigma_0^2/\hbar$.
That is, contrary to the classical description, quantum-mechanically
free expansion involves a stage of accelerated quantum motion.
This means that the classical limit will imply keeping this term
relatively small and, therefore, that QTs will be essentially parallel
to the classical one, $R_{cl}$, in agreement with Ehrenfest's theorem.

By inspecting (\ref{e12}), two timescales become apparent
\cite{sanz-cpl1}.
If $t \ll \tau_q$, the WP spreading is meaningless ($\sigma_t
\approx \sigma_0$) and QTs are all essentially
parallel to $R_{cl}$ (Ehrenfest's or classical criterion),
since $R_q(t) \approx R_{q,0} + R_{cl}(t)$.
In this case, the classical drift becomes the leading term.
As $t$ increases, if Eq.~(\ref{e13}) is expanded linearly in time, we
start to observe the action of the quantum component of the velocity.
This leads to an incipient accelerated motion, which at early stages
can be described according to the familiar expression from classical
mechanics
\begin{equation}
 R_q(t) \approx R_{q,0} + v_0 t + \frac{1}{2}\ a_q t^2 ,
 \label{e15}
\end{equation}
although the acceleration $a_q \equiv R_{q,0}/\tau_q^2$ here depends
on the initial position $R_{q,0}$ ---the larger the distance
$|R_{q,0} - R_{cl,0}|$, the larger the effects due to this quantum
acceleration.
As time increases asymptotically ($t \gg \tau_q$), we find $\sigma_t$
approaches a linear dependence on time and the WP curvature becomes
invariant under translations in time (this is the so-called Fraunhofer
regime in optics, where phases depend linearly on coordinates).
In this case, QTs go as
\begin{equation}
 R_q(t) \approx \left( v_0 + \frac{R_{q,0}}{\tau_q} \right) t ,
 \label{e16}
\end{equation}
i.e., the asymptotic motion is again uniform, but with the important
difference that the corresponding (constant) velocity having a
component proportional to the initial position of the trajectory.
This extra QM component is very important: in the
case of diffraction by periodic structures, for example, it is
precisely the one associated with the different diffraction channels
\cite{sanz-talbot} or with the transfer of crystalline momentum
\cite{sanz11,sanz12}.

In terms of the 2D problem, this means that at the first stages
of its evolution, $\phi_\gamma$ will undergo a kind of hyperbolic
expansion along the $R$-coordinate, independent of $r$ (where the
trajectories will remain essentially steady) and $\gamma$ (for any
angle we will observe a similar behavior initially).
Actually, in order to determine the leading dynamics from the initial
state (i.e., whether the spreading will dominate over the propagation
or vice versa), with the values given in Table~\ref{tabTDWP} we find
that the value of the ratio $v_0/v_s = 2k_0\sigma_0$ \cite{sanz-JPA}
is approximately 0.027.
This means that the WP (\ref{eq11}) will propagate along $R$ much
slower than it spreads.
Therefore, essentially at least half of it will move backwards and
will not contribute to the resonance (in other words, the associated
amount of reactants will not be reactive).
Once the onwards part of the WP (i.e., that spreading towards smaller
values of $R$) starts reaching the transition state region, the
approach considered here will not be valid any more (we will be in
the regime $t \ge t_{int}$).
In this case, it is also possible that, due to reflection or
interference \cite{sanz-bofill1,sanz-bofill2}, some additional portion
of reactants will also become inactive, bouncing backwards.


\section{Numerical simulations}
\label{sec3}


\subsection{The model}
\label{sec31}

\begin{figure}[!t]
 \begin{center}
  \includegraphics[width=7cm]{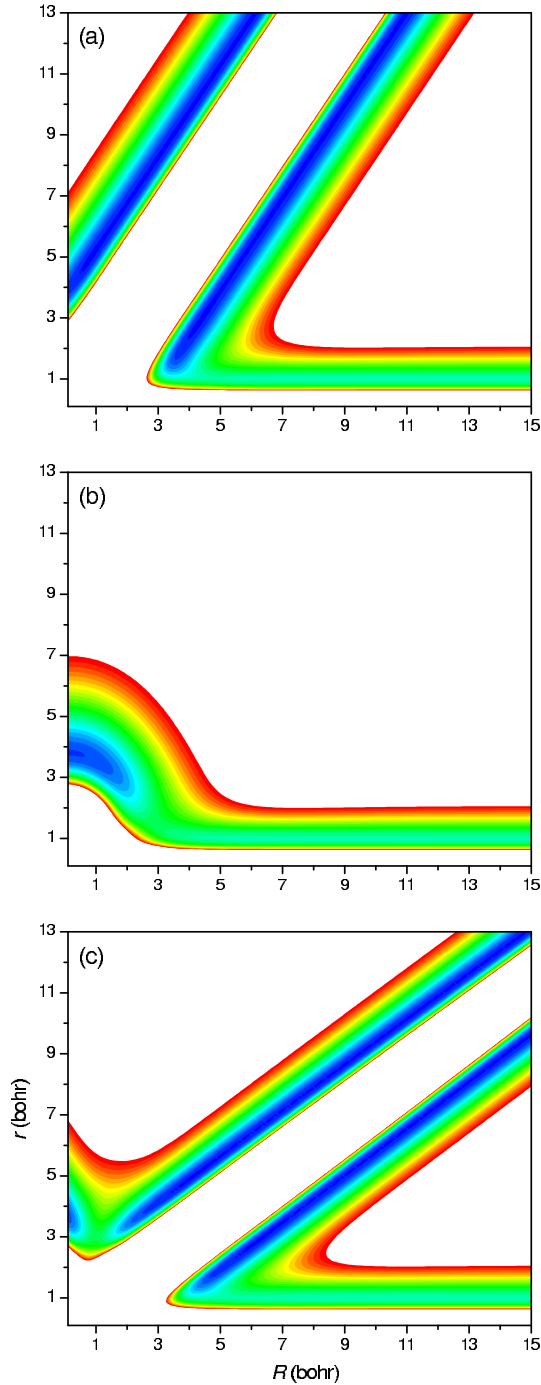}
  \caption{2D PES (slices) corresponding to the three angles selected
   here: (a) $\gamma = 0^\circ$, (b) $\gamma = 90^\circ$, and
   (c) $\gamma = 180^\circ$.
   Colored contours cover a range of energies between the minimum
   energy (deep blue), around $-1.5$~eV, and a maximum energy of
   3~eV (red).}
  \label{fig1}
 \end{center}
\end{figure}

In the ($R,r$) subspace, the SW PES \cite{SW:JCP96} employed here
to study the F+HD reaction is strongly dependent on the Jacobi
angular coordinate, $\gamma$, as can be seen in Fig.~\ref{fig1}, where
a contour-plot of this PES at three different angles is displayed.

In panel (a), for $\gamma = 0^\circ$, the entrance F+HD channel leads
smoothly to the FD+H product channel which exhibits two separated potential
minimum regions, a feature due to the possible atom permutations on the
corresponding diatom.
As $\gamma$ increases these two regions start to get closer and finally
merge on an unique minimum well at $\gamma = 90^\circ$ (see panel (b)).
If we keep increasing the angle, the PES starts to describe the other
possible product channel, FH+D.
Again, the presence of a significantly large barrier creates two unconnected
regions, with the separation getting at its highest for
$\gamma = 180^\circ$ (see panel (c)).
These projections of the PES at specific angles reveal that the connection
of the separated regions observed for $\gamma = 0^\circ$ and 180$^\circ$
is in fact only possible via the $\gamma \sim 90^\circ$ direction.
That means that, whereas this pathway is certainly possible for the 3D
evolution of the WP, restrictions are expected for the reduced 2D
dynamics of the QTs.

As mentioned above, the reduced 2D projection is extracted from the
3D WP at each time step $\delta t$ by fixing the angular coordinate,
$\gamma$.
It is worth noticing that the thus obtained evolution of the 2D WP onto
the ($R,r$) subspace will not be equivalent at all (unless the angular
degree of freedom is completely decoupled) to the evolution of a
strictly 2D WP along any of the potential surfaces displayed in
Fig.~\ref{fig1}.
On the contrary, due to the coupling to the angular coordinate, the
3D WP will lead the evolution of its 2D projection, avoiding motions
that one could observe otherwise.

\begin{figure}[!t]
 \begin{center}
  \includegraphics[width=7cm]{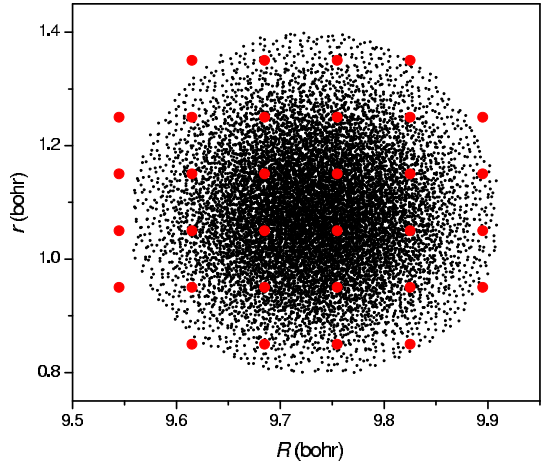}
  \caption{Monte-Carlo-like sampling of the initial conditions for the
   QTs using the initial probability density.
   The red bigger dots mark the position of the tracer QTs (see text for
   details).}
  \label{fig2}
 \end{center}
\end{figure}

The initial conditions for the corresponding QTs will be chosen taking
into account the initial 2D projected probability density at a specific
value $\gamma_0$ for the angle,
\begin{equation}
 \rho_{\gamma_0} = |\phi_{\gamma_0}(0)|^2
  = |G(R) \phi_{vj}(r) Y_{lj}^{JM}(\gamma_0)|^2 .
 \label{initdist}
\end{equation}
Although this initial-condition density distribution can be approximated by
a Gaussian function in both directions, we have implemented a Monte-Carlo-like
distribution based on defining a grid on it and populating each cell
according to the corresponding average weight.
This procedures renders a fair distribution, as can be seen in
Fig.~\ref{fig2}, which consists of about 12,800~QTs
when we choose a 100$\times$100 grid centered on $\rho_{\gamma_0}$ and
impose a maximum of 20 trajectories for the maximum value of
$\rho_{\gamma_0}$.
Apart from that, a less fine grid is also chosen to monitor the
evolution of relatively large samples of QTs.
The center of each cell of this grid is also chosen as the initial
condition for a QT, namely the tracer QT,
which will help us to understand the coarse grained
evolution of the initial WP.
In Fig.~\ref{fig2}, these tracer trajectories appear as red full
circles.

Once the initial conditions are set up, QTs are
obtained by integrating Eqs.~(\ref{dotR}) and (\ref{dotr}) with the value
of the projected 2D WP that we obtained from the propagation from $t$
to $t + \tau$ of the full 3D WP.
It is clear that those trajectories approaching the limits of the
entrance and exit channels may undergo the effects of the absorption.
In principle, one expects that the phase of the guiding wave function
will fade gradually and without changing much with respect to the
unabsorbed wave function.
In order to avoid eventual numerical problems,
the evolution of those trajectories that approach the vicinity of the
grid boundaries is stopped.
That is, each time a QT reaches a maximum value for $R$
and/or $r$, its evolution gets frozen.
Note that this procedure is physically reasonable, since any trajectory
moving backwards along the F+HD channel or forward along the FD+H or
FH+D channels will already be moving asymptotically and, therefore,
will not contribute anymore to the reaction dynamics.


\subsection{Analysis of WP dynamics and survival probabilities}
\label{sec32}

In order to monitor the formation of the transition state resonance
and its survival, as well as the passage to products, when working
within the time-dependent domain the most appropriate tool is to
define a survival or formation probability \cite{sanz-jcp-sars},
\begin{equation}
 \mathcal{P}_\Sigma (t) \equiv \int_\Sigma \rho_{\gamma_0}(t) dR dr .
 \label{eq-28}
\end{equation}
This quantity gives us the ``amount'' of probability inside a certain
region $\Sigma$ as a function of time.
Furthermore, its variation can be used as an indicator of the velocity
of the reaction under study.
In our case, we can consider the survival probability associated with
reactants, products and the transition state resonance.
This means that three $\Sigma$ regions have to be defined.
For example, looking at Fig.~\ref{fig1}(a), the transition state
resonance region, $\Sigma_{TSR}$, can be defined as the region enclosing
the ``elbow'' of the PES; the reactants region, $\Sigma_R$, as the
region enclosing only the entrance F+HD channel; and the products
region, $\Sigma_P$, as the region above $\Sigma_{TSR}$ (which would
include both exit channels).

It is worth mentioning that this approach differs from the spectral
quantization analysis performed by Skodje {\it et al.} \cite{SSMLDL:JCP00}.
In that work, the transition state was probed by explicitly setting the initial
WP at this region.
Here, the survival probabilities are defined in terms of a WP which initiates
its evolution from the reactant arrangement.

\begin{figure}[!ht]
 \begin{center}
  \includegraphics[width=7cm]{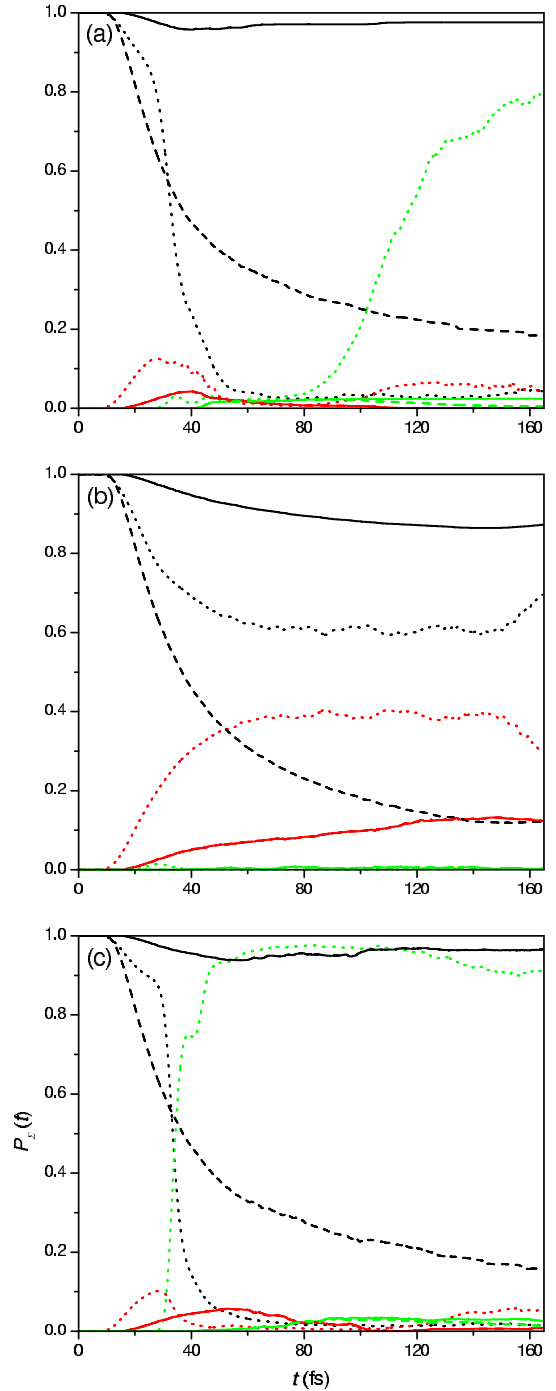}
  \caption{Reactants (black), transition-state (red) and products
   (green) probabilities as a function of time, obtained from the
   2D reduced WP (dotted line), QTs (solid lines) and QTs with
   absorption (dashed lines).
   From top to bottom, these probabilities refer to the three
   different angles considered here: (a) $\gamma_0 = 0^\circ$,
   (b) $\gamma_0 = 90^\circ$, and (c) $\gamma_0 = 180^\circ$.}
  \label{fig3}
 \end{center}
\end{figure}

In Fig.~\ref{fig3} we have plotted the three different probabilities
for the three values of $\gamma_0$ considered above (see dotted lines).
As can be noticed, the two extreme cases ($\gamma_0 = 0^\circ$ and
$\gamma_0 = 180^\circ$) present a certain resemblance: a fast decay of
the reactants probability between 30 and 50~fs and an equally fast
increase of the products probability, although this happens much
earlier and faster along the FH+D channel ($\gamma_0 = 180^\circ$).
The transition state resonance probability, though, acquires
relatively low values.
In the case of the T-shape configuration ($\gamma_0 = 90^\circ$), the
spatial limitations of the PES for the product channels
(see Fig.~\ref{fig1}(b)) make that all the
probability transferred goes to the intermediate resonance region.

\begin{figure}[!t]
 \begin{center}
  \includegraphics[width=7cm]{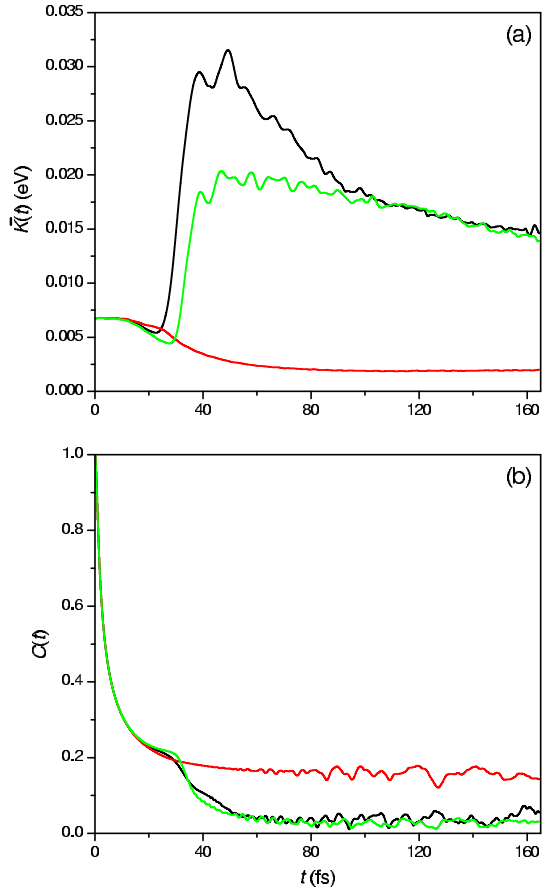}
  \caption{Average kinetic energy (a) and squared modulus of the time
   correlation function (b) associated with the reduced 2D wave
   function $\phi_{\gamma_0}$ as a function of time for the three the
   different angles considered here: $\gamma_0 = 0^\circ$ (black),
   $\gamma_0 = 90^\circ$ (red), and $\gamma_0 = 180^\circ$ (green).}
  \label{fig4}
 \end{center}
\end{figure}

The resemblance between the collinear configurations and their
difference with respect to the T-shape one can also be seen if
we compute the average kinetic energy associated with $\phi_{\gamma_0}$,
\begin{equation}
 \bar{K} = - \frac{\hbar^2}{2m}\
  \langle \phi_{\gamma_0}(t) | \left( \frac{\partial^2}{\partial R^2}
   + \frac{\partial^2}{\partial r^2} \right) | \phi_{\gamma_0}(t) \rangle ,
 \label{avkin}
\end{equation}
or the squared modulus of its time correlation function,
\begin{equation}
 C(t) = |\langle \phi_{\gamma_0}(0) | \phi_{\gamma_0}(t)\rangle |^2 .
 \label{corr}
\end{equation}
In Fig.~\ref{fig4} we have plotted these two quantities as a function
of time for the three angles considered.
As can be seen, in panel (a) we find that $\bar{K}$ increases quickly
and then decays more slowly and exponentially for $\gamma_0 = 0^\circ$
and 180$^\circ$, while it decreases and becomes almost constant for
$\gamma_0 = 90^\circ$.
Nonetheless, the maximum for $\gamma_0 = 180^\circ$ is less pronounced
than for $\gamma_0 = 0^\circ$, which is connected with the
the different dynamics expected for the two possible product arrangements.
The existence of a resonance for the F+HD $\to$ FH+D channel leads the
WP to stay longer at the spatially confined transition state region,
while the unperturbed evolution associated with a direct mechanism
that governs the dynamics for the F+HD $\to$ FD+H process.
This distinct behavior could account for the larger kinetic energy released
for this latter channel. Additional support to this hypothesis will be
discussed in the next sections.
Regarding $C(t)$, we find a similar trend for the two limiting cases,
with an abrupt decay around 30 fs that does not appear for the
T-shape configuration.

A similar analysis of survival probabilities can be now carried out
in terms of QTs.
In this case, the meaning of survival probability becomes
closely related,
as in classical statistical treatments, to
the number of trajectories $N_\Sigma$ (each associated with a system
configuration) that remain inside the corresponding region at a
given time in relation to the total number of trajectories $N$
considered in the statistics \cite{sanz-jcp-sars},
\begin{equation}
 \mathcal{W}(t) \equiv \frac{N_\Sigma (t)}{N} .
 \label{eq-29}
\end{equation}
In principle, if the dimensionality of the trajectory-based
calculation would be the same as the WP one, $\mathcal{W}$ would
approach $\mathcal{P}(t)$ as $N \to \infty$ (given an initial
sampling according to $\rho_0$).
Since this is not the case here for the two simulations performed,
$\mathcal{W}$ will give us information about the
2D dynamics, which may differ importantly from the full 3D WP
simulation.
This is precisely what we observe in Fig.~\ref{fig3} with solid
line: a very small portion of the initial configurations sampled
will contribute to the resonance and to products, since most of
the probability will flow backwards (see next Section).
Of course, as some probability is being absorbed,
we can also compute $\mathcal{W}$ in terms of a {\it variable}
total number of trajectories, which gives us the total amount of
trajectories that are still inside certain boundaries (see
previous Section).
However, the corresponding results (dashed lines in Fig.~\ref{fig3})
show that the probability mainly concentrates along the reactants
regions (indeed, an important portion of it flows backwards, towards
large $R$), with no significant contribution from the resonance-forming
or the appearance of products processes.


\subsection{Analysis of quantum trajectories}
\label{sec33}

\begin{figure*}[!t]
 \begin{center}
  \includegraphics[width=16cm]{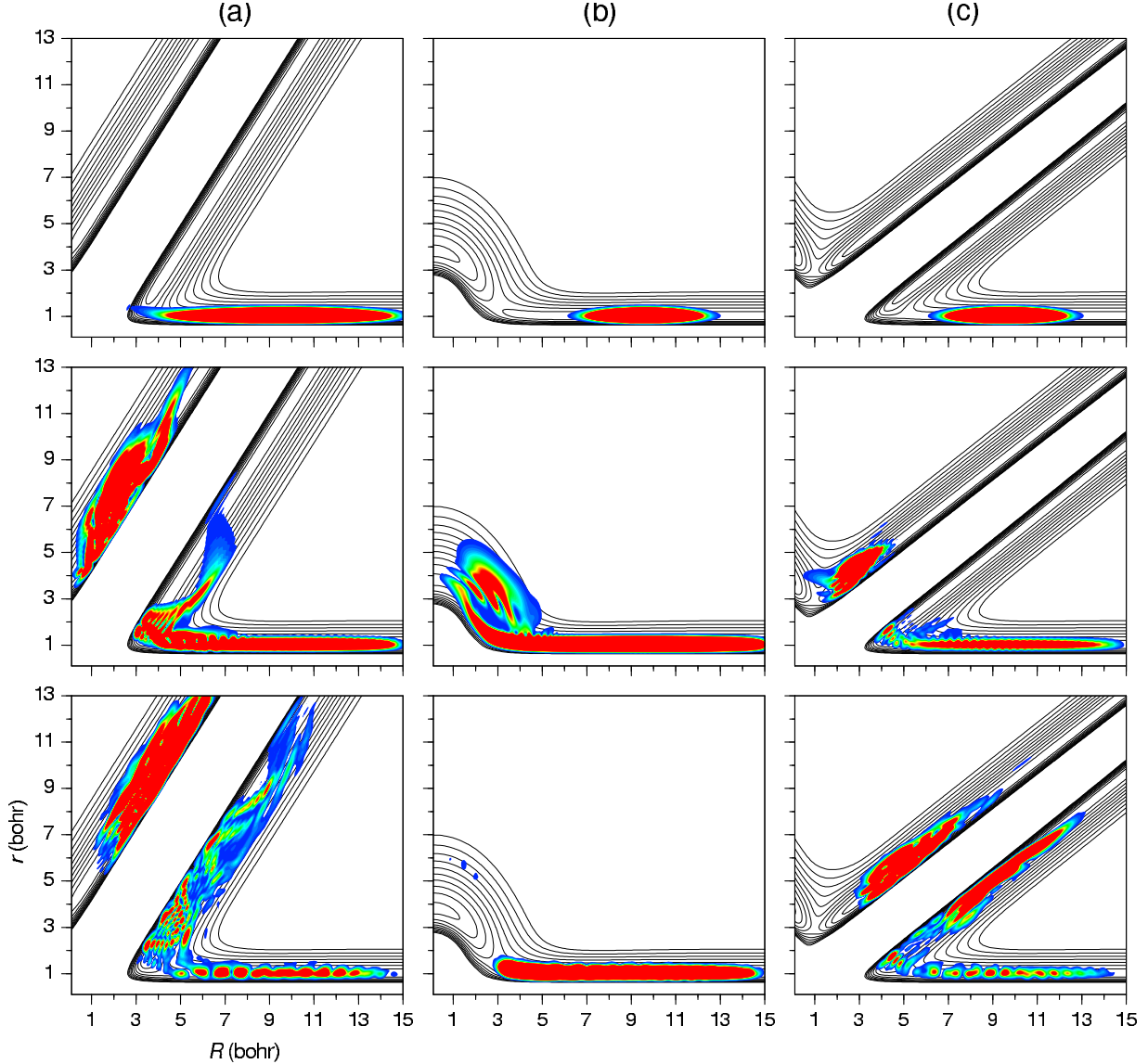}
  \caption{Snapshots illustrating the time-evolution of the reduced
   2D WPs corresponding to the three different angles considered here:
   (a) $\gamma_0 = 0^\circ$, (b) $\gamma_0 = 90^\circ$, and
   (c) $\gamma_0 = 180^\circ$.
   From top to bottom, the times associated with each frame of the
   three cases considered are: (a)~13.2~fs, 39.5~fs, and 72.4~fs;
   (b) 6.6~fs, 26.3~fs, and 65.8~fs; (c) 6.6~fs, 39.5~fs, and 65.8~fs.}
  \label{fig5}
 \end{center}
\end{figure*}

In the previous Section we have analyzed ``ensemble'' or statistical
properties, without paying any attention to the way how the probability
flows given a certain angular projection.
Thus, let us consider now the evolution of the QTs,
in particular the tracer QTs for each one of the angular
configurations given above.
In this regard, it might be insightful first to consider a few
snapshots of the evolution of the reduced 2D WPs for each angle.
This is illustrated in Fig.~\ref{fig5}.

The evolution for the $\gamma_0 = 0^\circ$ case ((a) panels on the left of
Fig.~\ref{fig5}) shows the WP extending on both branches with a rich nodal
structure around the intermediate range.
These features though do not seem to remain as the propagation time evolves,
thus merely indicating the existence of the expected direct process for
the F+HD $\to$ FD+H case.
For $\gamma_0 = 90^\circ$ ((b) middle panels of Fig.~\ref{fig5}), the required
bridge between the separated branches shown by the PES at some other angular
directions, the WP is restricted to a very localized area at the intermediate
region.
After spending some time there, a significant component of the WP moves backward into
the reactant channel.
The situation at $\gamma_0 = 180^\circ$ ((c) panels on the right of
Fig.~\ref{fig5}) clearly reveals the presence of a resonance at the
transition state in the FH+D product channel.
As the WP reaches the intermediate region a stable 3-node structure is formed,
governing the passage to both the upper and lower structures of the exit
channels.

\begin{figure}[!ht]
 \begin{center}
  \includegraphics[width=6.75cm]{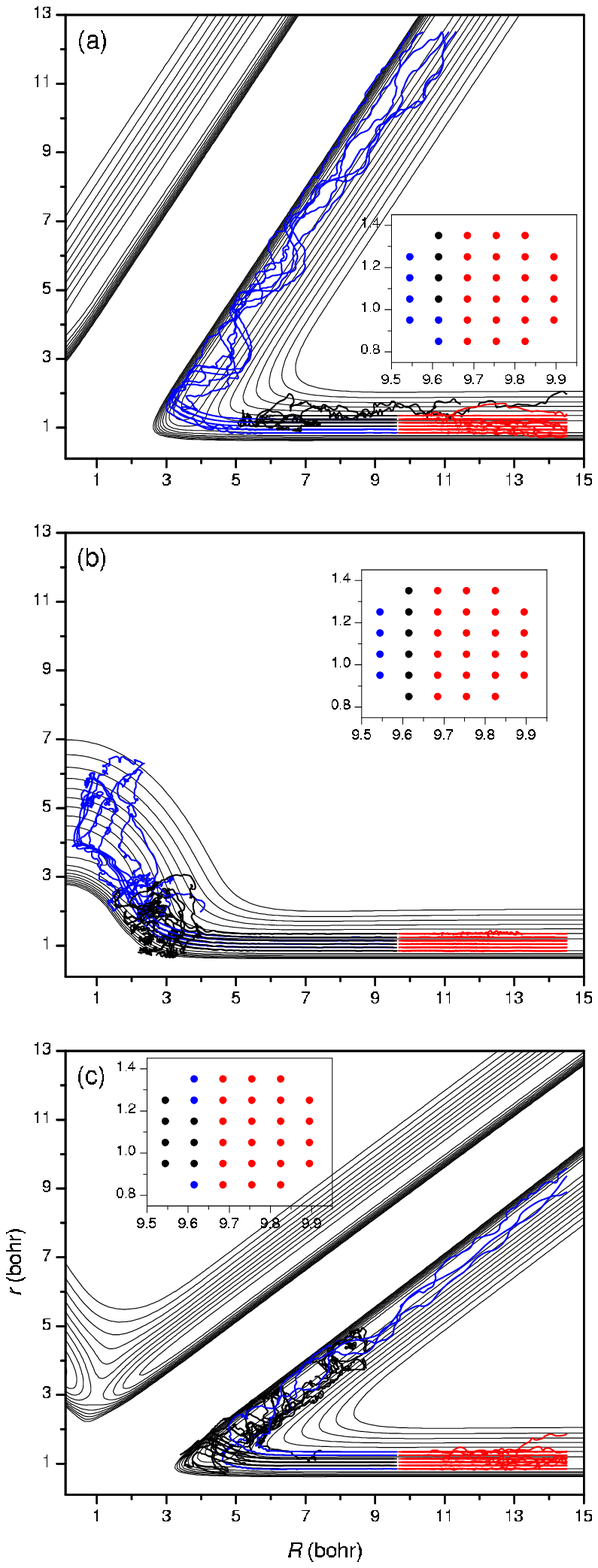}
  \caption{Tracer QTs corresponding to the three
   different angles considered here: (a) $\gamma_0 = 0^\circ$,
   (b) $\gamma_0 = 90^\circ$, and (c) $\gamma_0 = 180^\circ$.
   Non-reactive trajectories are displayed with red color, while
   reactive ones in (a) and (c) are in blue, as well as those
   transiently passing to the second channel in (b); with black,
   tracer QTs that reach the transition state
   region and then are scattered back.
   In order to facilitate the identification of the initial
   conditions leading to each type of motion, they have been
   plotted in the insets with the corresponding colors.}
  \label{fig6}
 \end{center}
\end{figure}

In Fig.~\ref{fig6} we present the QT counterpart of
Fig.~\ref{fig5}.
The first remarkable feature, as indicated above, is that more than a
half of the initial conditions lead to backward motions, this being
the cause of the high probability in the reactants region.
This justifies the fact of using filtering techniques in order to
detect transition state resonances, for the corresponding probability
is so small compared to the reactants probability that it would be
difficult to detect them otherwise.
Now, if we compare the three angular configurations, we readily
notice that the number of tracer trajectories remaining in the
resonance region (or in a neighborhood) is much larger in the case of
$\gamma_0 = 180^\circ$ (see panel (c))than for $\gamma_0 = 0^\circ$
(panel (a)), where all the trajectories display a direct passage to
products.
This result is consistent with the features of the resonance observed
in the TDWP calculation for $\gamma_0 = 180^\circ$
discussed above (see Fig.~\ref{fig5}).
Moreover, the behavior of the QTs in this particular region, where
it is spatially confined, would also support a smaller amount of
average kinetic energy as compared with the FD+H channel as commented
in the previous section.

On the other hand, in panel (b) we notice that, for the T-shape
configuration ($\gamma_0 = 90^\circ$), trajectories concentrate for a
time around two different regions, which essentially coincide with
the separated areas of the FD+H and FH+D channels (see
Fig.~\ref{fig1}).
Furthermore, it is also worth stressing the fact that, since the
dynamics is restricted to a 2D plane, the QTs
starting in the F+HD channel will only be able to
just one of the available regions for the
FD+H channel (see panel (a)) or the FH+D one (see panel (c)).
The component of the fully 3D WP which explores the upper branch of the
PES contours does not manage to carry the corresponding 2D QTs.
The trajectory seems to perceive it but not populate it.
That is, it is a sort of ``empty wave'' \cite{holland-bk}.
In order to make active this part of the wave function, one should
consider that initially there is some set of steady initial conditions
resting in the associated channel.
Then, as this projection of the 3D wave function would start emerging,
those initial conditions would start to evolve accordingly.

\begin{figure}[!t]
 \begin{center}
  \includegraphics[width=7cm]{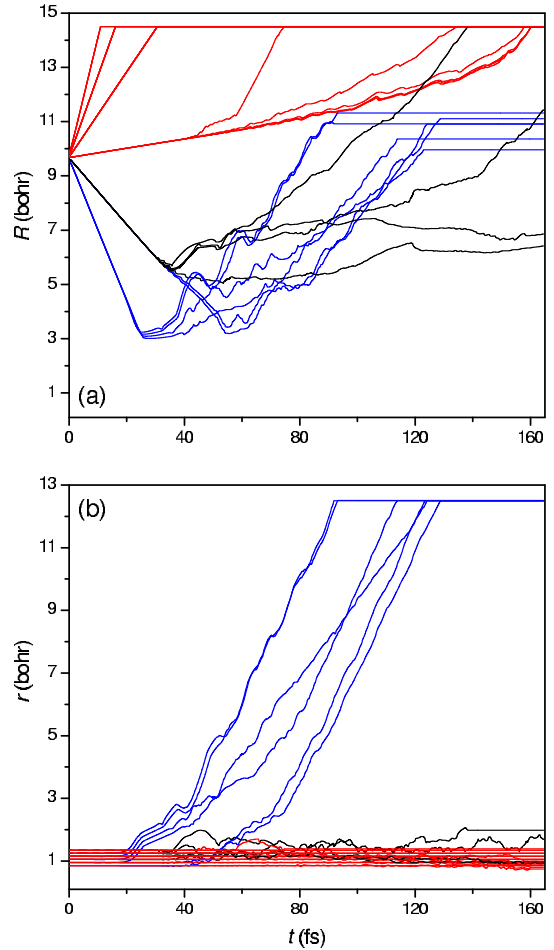}
  \caption{Time-dependence of the $R$ (a) and $r$ (b) degrees of freedom
   of the tracer QTs corresponding to $\gamma_0 = 0^\circ$.
   Colors are as in Fig.~\ref{fig6}.}
  \label{fig7}
 \end{center}
\end{figure}

\begin{figure}[!t]
 \begin{center}
  \includegraphics[width=7cm]{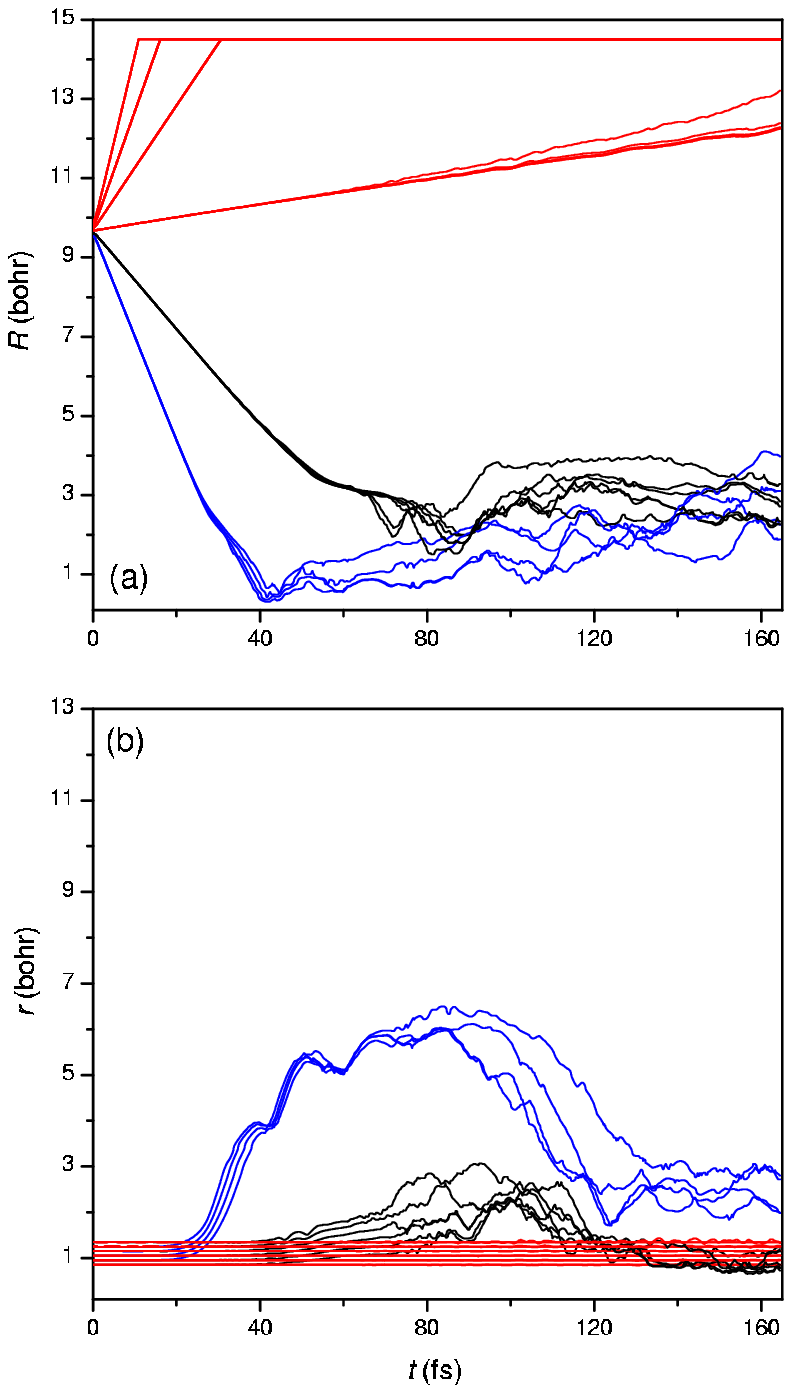}
  \caption{Time-dependence of the $R$ (a) and $r$ (b) degrees of freedom
   of the tracer QTs corresponding to $\gamma_0 = 90^\circ$.
   Colors are as in Fig.~\ref{fig6}.}
  \label{fig8}
 \end{center}
\end{figure}

\begin{figure}[!t]
 \begin{center}
  \includegraphics[width=7cm]{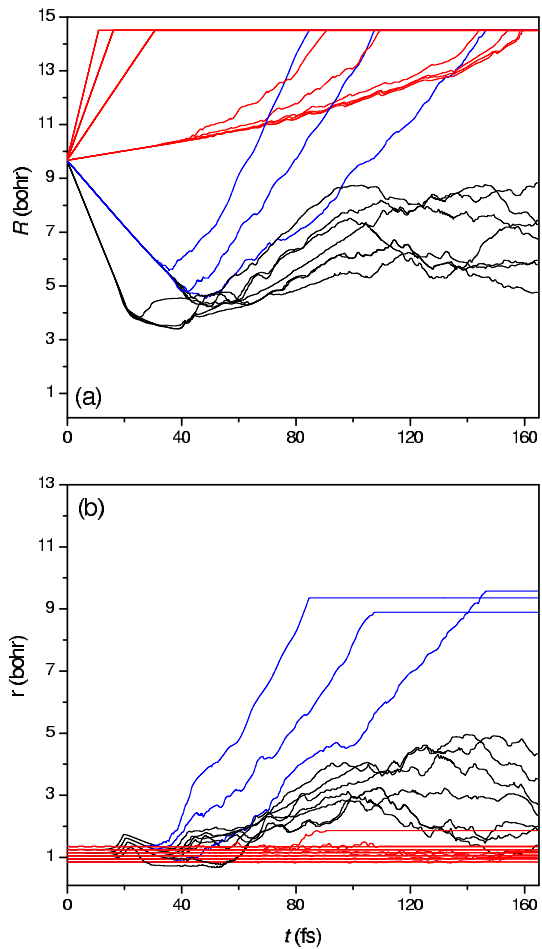}
  \caption{Time-dependence of the $R$ (a) and $r$ (b) degrees of freedom
   of the tracer QTs corresponding to $\gamma_0 = 180^\circ$.
   Colors are as in Fig.~\ref{fig6}.}
  \label{fig9}
 \end{center}
\end{figure}

A complementary picture of the time-evolution of the QTs
arises when we look at each degree of freedom separately as a function of
time.
The corresponding plots are presented in Figs.~\ref{fig7}, \ref{fig8},
and \ref{fig9}, respectively.
Taking into account this viewpoint, first we notice that for
$\gamma_0 = 0^\circ$ the $R$-component behaves according to the
prescription for a Gaussian WP at early stages of the evolution,
as indicated in Section~\ref{sec23},
while the $r$-component is basically steady (this behavior can also be observed
for the other two angular configurations).
Then, we find the three well-defined sets of QTs: those moving
backwards (red), those moving backwards after reaching the ``elbow''
of the reduced PES (black), and those passing to products (blue).
If we go to $\gamma_0 = 90^\circ$ (see Fig.~\ref{fig8}), we readily notice the
appearance of two groups of trajectories, one
exploring the beginning of the upper region observed in both product channels
(blue) and another one into the lower structure (black).
However, due to the constraint of 2D motion, none of them can indeed penetrate
into any of the channels and the trajectories are bounced backwards.
Finally, in the case of $\gamma_0 = 180^\circ$ (see Fig.~\ref{fig9}), we observe
a behavior analogous to the one seen for $\gamma_0 = 0^\circ$, although now only
a smaller portion of the QTs passes to products, the remaining
ones keeping trapped around the corresponding PES ``elbow'' (see Fig.~\ref{fig1}(a)).

\section{Conclusions}
\label{sec4}

The dynamics of the F+HD reaction for $J=0$ has been investigated by
means of a TDWP calculation combined with a QT approach, where the
latter was based on computing the trajectories associated with 2D WPs
arisen from the projections of the full 3D calculations at given values
of the angular coordinate.
The evolution of the trajectories has been found to be limited when
an explicit angular correlation is required.
However, the distinct reaction mechanism for each product channel is
successfully described.
A clear indication of the resonance-mediated pathway for the FH
forming arrangement thus becomes evident within the QT analysis, whereas
the 2D evolution of the trajectories for the FD+H product channel is
consistent with the observed direct reaction mechanism.

As mentioned in the Introduction, in this work we have focused on the
reduced radial subspace in order to avoid the complexity of the full 3D
QT dynamics.
For the same reason, we have chosen the case $J = 0$.
This does not mean that the method is limited either theoretically or
computationally.
Rather it constitutes a firs step in the development of a more general
QT methodology aimed at analyzing resonance formation processes.
Notice that, by proceeding in this way, one gains an insight on the dynamics
that can be later on used to understand in depth the full 3D dynamics.
In this regard, if the 2D reduced QT dynamics provides us with interesting
information about the flux of probability in these processes for a fixed
angle, the 3D dynamics will allow us to understand how this flux is
transferred from among different rotational populations.
Nevertheless, work in the direction of recreating the full 3D dynamics
is currently under development.


\section*{Acknowledgements}

It is a pleasure for the authors to dedicate this work to Prof.\ Gerardo
Delgado-Barrio, an inspiring scientist for all of us.

Support from the Ministerio de Ciencia e Innovaci\'{o}n (Spain) under
Projects FIS2010-18132 and FIS2010-22082 is acknowledged.
A.S. Sanz and D. L\'opez-Dur\'an respectively thank the Ministerio de
Ciencia e Innovaci\'on for a ``Ram\'on y Cajal'' Research Fellowship
and the Consejo Superior de Investigaciones Cient{\'\i}ficas for a
``JAE-Doc'' contract.







\begin{thebibliography}{99}

\bibitem{bohm}
 D. Bohm, Phys. Rev. 85 (1952) 166; 85 (1952) 180.

\bibitem{holland-bk}
 P. R. Holland, The Quantum Theory of Motion,
 Cambridge University Press, Cambridge, 1993.

\bibitem{wyatt-bk}
 R. E. Wyatt, Quantum Dynamics with Trajectories,
 Springer, New York, 2005.

\bibitem{wyatt1}
 C. L. Lopreore, R. E. Wyatt, Phys. Rev. Lett. 82 (1999) 5190.

\bibitem{wyatt2}
 C. L. Lopreore, R. E. Wyatt, Chem. Phys. Lett. 325 (2000) 73.

\bibitem{sanz-ssr}
 R. Guantes, A. S. Sanz, J. Margalef-Roig, S. Miret-Art\'es,
 Surf. Sci. Rep. 53 (2004) 199.

\bibitem{sanz-pr}
 A. S. Sanz, S. Miret-Art\'es, Phys. Rep. 451 (2007) 37.

\bibitem{madelung}
 E. Madelung, Z. Phys. 40 (1926) 322.

\bibitem{birula1}
 I. Bialynicki-Birula and Z. Bialynicka-Birula,
 Phys. Rev. D 3 (1971) 2410.

\bibitem{birula2}
 I. Bialynicki-Birula, M. Cieplak, J. Kaminski,
 Theory of Quanta, Oxford University Press, Oxford, 1992.

\bibitem{hirsch1}
 J. O. Hirschfelder, A. C. Christoph, W. E. Palke,
 J. Chem. Phys. 61 (1974) 5435.

\bibitem{hirsch2}
 J. O. Hirschfelder, C. J. Goebel, L. W. Bruch,
 J. Chem. Phys. 61 (1974) 5456.

\bibitem{hirsch3}
 J. O. Hirschfelder, K. T. Tang, J. Chem. Phys. 64 (1976) 760.

\bibitem{hirsch4}
 J. O. Hirschfelder, K. T. Tang, J. Chem. Phys. 65 (1976) 470.

\bibitem{MC-W1}
 E. A. McCullough, R. E. Wyatt, J. Chem. Phys. 51 (1969) 1253.

\bibitem{MC-W2}
 E. A. McCullough, R. E. Wyatt, J. Chem. Phys. 54 (1971) 3578.

\bibitem{MC-W3}
 E. A. McCullough, R. E. Wyatt, J. Chem. Phys. 54 (1971) 3592.

\bibitem{sanz-bofill1}
 A. S. Sanz, X. Gim\'enez, J. M. Bofill, S. Miret-Art\'es,
 Chem. Phys. Lett. 478 (2009) 89.

\bibitem{sanz-bofill2}
 A. S. Sanz, X. Gim\'enez, J. M. Bofill, S. Miret-Art\'es,
 Chem. Phys. Lett. 488 (2010) 235.

\bibitem{muellerbrown}
 K. M\"uller, L. D. Brown, Theor. Chim. Acta 53 (1979) 75.

\bibitem{as:bader}
 R. F. W. Bader, J. Chem. Phys. 73 (1980) 2871.

\bibitem{as:gomes-1}
 J. A. N. F. Gomes, J. Chem. Phys. 78 (1983) 3133.

\bibitem{as:gomes-2}
 J. A. N. F. Gomes, J. Chem. Phys. 78 (1983) 4585.

\bibitem{as:lazzeretti-1}
 P. Lazzeretti, Prog. Nuc. Mag. Res. Spect. 36 (2000) 1.

\bibitem{as:lazzeretti-2}
 S. Pelloni, F. Faglioni, R. Zanasi, P. Lazzeretti,
 Phys. Rev. A 74 (2006) 012506.

\bibitem{as:lazzeretti-3}
 S. Pelloni, P. Lazzeretti, R. Zanasi,
 J. Phys. Chem. A 111 (2007) 3110.

\bibitem{as:lazzeretti-4}
 S. Pelloni, P. Lazzeretti, R. Zanasi,
 Theor. Chem. Acc. 123 (2009) 353.

\bibitem{as:lazzeretti-5}
 S. Pelloni, P. Lazzeretti,
 J. Phys. Chem. A 112 (2008) 5175.

\bibitem{as:lazzeretti-6}
 S. Pelloni, P. Lazzeretti,
 J. Chem. Phys. 128 (2008) 194305.

\bibitem{as:lazzeretti-7}
 S. Pelloni, P. Lazzeretti,
 Chem. Phys. 356 (2009) 153.

\bibitem{as:lazzeretti-8}
 I. Garc\'{\i}a Cuesta, A. S\'anchez de Mer\'as, S. Pelloni,
 P. Lazzeretti, J. Comput. Chem. 30 (2009) 551.

\bibitem{SSMLDL:JCP00}
 R. T. Skodje, D. Skouteris, D. E. Manolopoulos, S.-H. Lee, F. Dong, K. Liu
 J. Chem. Phys. 112 (2000) 4536.

\bibitem{SSMLDL:PRL00}
 R. T. Skodje, S. Skouteris, D. E. Manolopoulos, S.-H. Lee, F. Dong, K. Liu
 Phys. Rev. Lett. 85 (2000) 1206.

\bibitem{SW:JCP96}
 K. Stark, H.-J. Werner, J. Chem. Phys. 104 (1996) 6515.

\bibitem{ABHSSW:JCP95}
 F. J. Aoiz, L. Ba\~nares, V. J. Herrero, V. S\'aez R\'abanos, K. Stark,
 H.-J. Werner, J. Chem. Phys. 102 (1995) 9248.

\bibitem{ABHSSTW:CPL96}
 F. J. Aoiz, L. Ba\~nares, V. J. Herrero, V. S\'aez R\'abanos, K. Stark,
 I. Tanarro, H.-J. Werner, Chem. Phys. Lett. 262 (1996) 75.

\bibitem{GRM:JCP04}
 T. Gonz\'alez-Lezana, E. J. Rackham, D. E. Manolopoulos,
 J. Chem. Phys. 120 (2004) 2247.

\bibitem{FF:JCP83}
 M. D. Feit, J. A. Fleck, Jr., J. Chem. Phys. 78 (1983) 301.

\bibitem{M:JCP02}
 D. E. Manolopoulos, J. Chem. Phys. 117 (2002) 9552.

\bibitem{borondo1}
 A. S. Sanz, F. Borondo, Eur. Phys. J. D 44 (2007) 319.

\bibitem{borondo2}
 A. S. Sanz, F. Borondo, Chem. Phys. Lett. 478 (2009) 301.

\bibitem{sanz-cpl1}
 A. S. Sanz, S. Miret-Art\'es,
 Chem. Phys. Lett. 445 (2007) 350.

\bibitem{sanz-talbot}
 A. S. Sanz, S. Miret-Art\'es,
 J. Chem. Phys. 126 (2007) 234106.

\bibitem{sanz11}
 A. S. Sanz, F. Borondo, S. Miret-Art\'es,
 Phys. Rev. B 61 (2000) 7743.

\bibitem{sanz12}
 A. S. Sanz, F. Borondo, and S. Miret-Art\'es,
 Europhys. Lett. 55 (2001) 303.

\bibitem{sanz-JPA}
 A.S. Sanz, S. Miret-Art\'es, J. Phys. A 41 (2008) 435303.

\bibitem{sanz-jcp-sars}
 A.S. Sanz, S. Miret-Art\'es, J. Chem. Phys. 122 (2005) 014702.

\end{thebibliography}
\end{document}